\documentclass[11pt,a4paper]{article}

\usepackage[margin=1in]{geometry}
\usepackage[T1]{fontenc}
\usepackage{lmodern}
\usepackage{microtype}
\usepackage{amsmath,amssymb}
\usepackage{graphicx}
\usepackage{booktabs}
\usepackage{array}
\usepackage{tabularx}
\usepackage{ragged2e}
\usepackage{xcolor}
\usepackage{hyperref}
\usepackage{natbib}
\usepackage{listings}

\hypersetup{
    colorlinks=true,
    linkcolor=blue,
    citecolor=blue,
    urlcolor=blue,
    pdftitle={Schema-Validated LLM Workflows: A Benchmark Protocol and Comparative Evaluation Across Six Python Implementations},
    pdfauthor={Sidney Shapiro, Sanad Al-Khatib, Aurora Pinz\'{o}n Arzola, Emiliano Sebastian Gonzalez Venegas, and Daniel Pearson},
    pdfsubject={Benchmark protocol and live-API pilot for schema-validated LLM business workflows},
    pdfkeywords={large language models, structured output, runtime validation, PydanticAI, agent frameworks, business workflows},
    pdflang={en-CA},
}

\lstdefinestyle{pythonstyle}{
    language=Python,
    basicstyle=\small\ttfamily,
    keywordstyle=\color{blue},
    commentstyle=\color{gray},
    stringstyle=\color{teal},
    breaklines=true,
    columns=fullflexible,
    keepspaces=true,
    showstringspaces=false,
    frame=single,
    framerule=0.4pt,
}

\lstdefinestyle{yamlstyle}{
    basicstyle=\small\ttfamily,
    breaklines=true,
    columns=fullflexible,
    frame=single,
    framerule=0.4pt,
}

\newcolumntype{P}[1]{>{\RaggedRight\arraybackslash}p{#1}}
\newcolumntype{Y}{>{\RaggedRight\arraybackslash}X}

\title{Schema-Validated LLM Workflows: A Benchmark Protocol and Comparative Evaluation Across Six Python Implementations}

\author{
Sidney Shapiro$^{1}$, Sanad Al-Khatib$^{2}$, Aurora Pinz\'{o}n Arzola$^{3}$,\\
Emiliano Sebastian Gonzalez Venegas$^{4}$, and Daniel Pearson$^{1}$\\[0.6em]
\small $^{1}$Dhillon School of Business, University of Lethbridge, Canada\\
\small $^{2}$Al Hussein Technical University, Jordan\\
\small $^{3}$Universidad de Guanajuato, Mexico\\
\small $^{4}$Universidad de Guadalajara, Mexico
}

\date{}

\begin{document}

\maketitle

\begin{abstract}
Large language models are increasingly used to extract records, generate SQL, summarize reports, answer policy questions, and route operational work. At these software boundaries, parseable output is necessary but does not establish task correctness. We present a reproducible benchmark protocol, a public executable artifact, and a confirmatory live-API evaluation that separate structural validity from task-specific correctness across five business-workflow tasks and six Python implementation conditions: PydanticAI, LangGraph, DSPy, CrewAI, Agno, and a framework-free SDK reference.

We hypothesize that (H1)~schema validity and task quality are separable, (H2)~structured-output binding paths are heterogeneous under a shared model and lowest-common-denominator contract, and (H3)~condition$\times$task interactions concentrate failures rather than shifting all tasks equally. Confirmatory analysis uses distinct task instances (100 extraction and 50 per other task, with three nested replicates) under a frozen analysis plan, yielding a 5{,}400-row primary matrix with equivalence margins and Holm-adjusted contrasts against the SDK reference.

In the eligible confirmatory run, overall schema validity was 0.684 and mean automated quality was 0.638. PydanticAI, LangGraph, and DSPy retained near-ceiling schema validity with mean quality $0.91$--$0.94$, while the plain SDK, Agno, and CrewAI conditions were substantially lower in this package and model configuration. Schema-valid rows still produced many task-score flags, especially in extraction and compliance QA, and quality residuals show strong condition$\times$task interactions (including a workflow collapse for CrewAI). Labeled ablations address retry budgets and feature-using mechanisms; primary contrasts remain system-configuration findings, not causal attributions to marketed library features.

\end{abstract}

\section{Introduction}
\label{sec:introduction}

Business workflows are a natural deployment target for large language models (LLMs). Analysts and operations teams use them to extract fields from documents, translate questions into SQL, summarize evidence, answer policy questions, and route requests through tools. These outputs are rarely consumed only as prose: they are parsed, stored, joined with enterprise data, or used to trigger actions. A fluent response that omits a field can break a pipeline, while a well-formed object can still contain the wrong amount, query, citation, or decision.

Reliable deployment therefore requires several distinct controls. Provider-native structured output constrains the generated representation. Application validation checks that the returned value satisfies a runtime schema and business rules. Framework mechanisms can add retries, typed dependencies, tools, state, and trace capture. Task-specific evaluation then checks whether the accepted value is correct for its source and intended use. Conflating these layers gives schema conformance credit for factual accuracy or gives a framework credit for behavior supplied by the provider API.

This paper's primary contribution is a \emph{protocol} that keeps those layers separable under matched business-workflow inputs, together with a confirmatory evaluation of six Python implementation conditions. We use \emph{schema-validated} to mean that an output has passed a declared runtime contract. We use \emph{structural validity} for parseability and schema conformance, and \emph{task-specific correctness} for agreement with source facts, SQL results, citations, or workflow ground truth. These definitions are intentionally narrower than end-to-end ``type safety'': runtime validation can reject malformed values, but it cannot prove factual or procedural correctness. We also use \emph{implementation condition}, rather than \emph{framework}, when referring collectively to PydanticAI, LangGraph, DSPy, CrewAI, Agno, and the framework-free SDK reference.

\paragraph{Working hypotheses.}
We test three claims that follow from that separation. First, structural validity and task correctness are separable: schema-valid objects can still fail deterministic task scorers (H1). Second, under a shared model and lowest-common-denominator structured-output contract, implementation binding paths are heterogeneous relative to an SDK baseline (H2). Third, gaps are interactive rather than purely additive: some tasks compress condition differences while others expose adapter failures (H3). Mechanism attribution---whether validators, retries, typed dependencies, or orchestration cause those gaps---is reserved for labeled ablations (RQ4).

Figure~\ref{fig:evaluation-layers} (Section~\ref{sec:background}) summarizes the evaluation path: task input $\rightarrow$ provider constraint and implementation condition $\rightarrow$ shared record and dual scoring (structural and task-specific). Existing structured-output benchmarks distinguish schema adherence from value correctness, including recent evidence that constraints can raise validity while increasing wrong-valid-schema answers \citep{geng2025jsonschemabench,yang2025structeval,singh2026sob,ray2026constrainttax}. Agent benchmarks emphasize tools, state, cost, and traces, with growing attention to full paths and deterministic validators rather than final-answer heuristics alone \citep{liu2024agentbench,ma2024agentboard,michelakis2025core,gurram2026toolvalidator}. What remains difficult is a matched comparison that connects both perspectives to common business-workflow tasks and preserves enough evidence to attribute failures.

The confirmatory design commits a primary matrix of six conditions, three nested replicates, and distinct instances ($100+50+50+50+50$), for 5{,}400 execution rows. Effective sample size is the number of distinct instances, not the row total. Equivalence margins, Holm multiplicity against the SDK reference, and simulation-based power are frozen in \texttt{config/analysis\_plan.json} before eligible live runs.

Pydantic is a common Python runtime-validation layer for typed data models and JSON Schema generation \citep{pydantic2026}. PydanticAI extends that model to LLM applications through structured outputs, typed dependencies, tools, validators, retries, and observability \citep{pydanticai2026,pydanticaiOutput2026}. Other libraries emphasize different mechanisms: LangGraph exposes stateful orchestration \citep{langgraph2026}, DSPy optimizes language-model programs against metrics \citep{khattab2023dspy}, CrewAI organizes role- and task-based execution \citep{crewai2026}, and Agno provides a compact typed-agent interface \citep{agno2026}. Provider APIs can now enforce JSON Schemas directly, so a plain SDK reference is necessary to measure how much additional behavior the application layer contributes \citep{openaiStructured2024}.

The paper makes three contributions:
\begin{enumerate}
\item A controlled benchmark protocol for five business-workflow tasks that separates structural validity, task-specific correctness, operational cost, and engineering burden, with harness-boundary revalidation and attempt-level provenance.
\item An executable comparison artifact with shared schemas and prompts, normalized trial records, framework adapters, hardened scorers, audit artifacts, frozen power/equivalence settings, and pre-specified ablation and stress conditions, released at \url{https://github.com/ULDataProfessor/schema-validated-llm-workflows}.
\item A confirmatory 5{,}400-row evaluation of six implementation conditions on distinct, label-derivable instances, testing separability, binding-path heterogeneity, and condition$\times$task interaction under a frozen analysis plan.
\end{enumerate}

The remainder of the paper defines the reliability layers and related evidence, states the research questions and hypotheses, describes the benchmark and frozen analysis plan, reports confirmatory results and ablations, and discusses technical and business implications. The final sections consolidate validity threats, artifact and ethical considerations, and the claim boundary.

\section{Reliability Layers}
\label{sec:background}

Pydantic turns Python annotations into runtime data contracts. A model defines fields, validation rules, serialization behavior, and a JSON Schema that can be shared with an external system \citep{pydantic2026}. For an LLM workflow, this contract governs whether application code accepts a returned value. It does not establish that the value is supported by a source document or appropriate for the business decision.

Provider-native structured-output modes move part of this control into generation. JSON Schema modes and tool interfaces can constrain responses to a declared representation, reducing malformed JSON and missing-field failures \citep{openaiStructured2024,openaiStructuredDocs2026,anthropicStructured2026}. Because the provider may already deliver ceiling-level conformance for a narrow schema, any application-framework comparison must keep provider enforcement visible as part of the treatment.

PydanticAI integrates Pydantic-backed outputs with dependencies, tools, output validators, retry behavior, and observability hooks \citep{pydanticai2026,pydanticaiOutput2026,pydanticaiDeps2026,pydanticaiTools2026}. The relevant empirical question is not whether these features exist, but whether exercising them changes an observed failure: for example, whether an output validator rejects unsafe SQL, a typed dependency makes a policy version auditable, or retry feedback repairs an otherwise rejected object. A condition that uses only provider-native output tests a much narrower mechanism.

\begin{figure}[t]
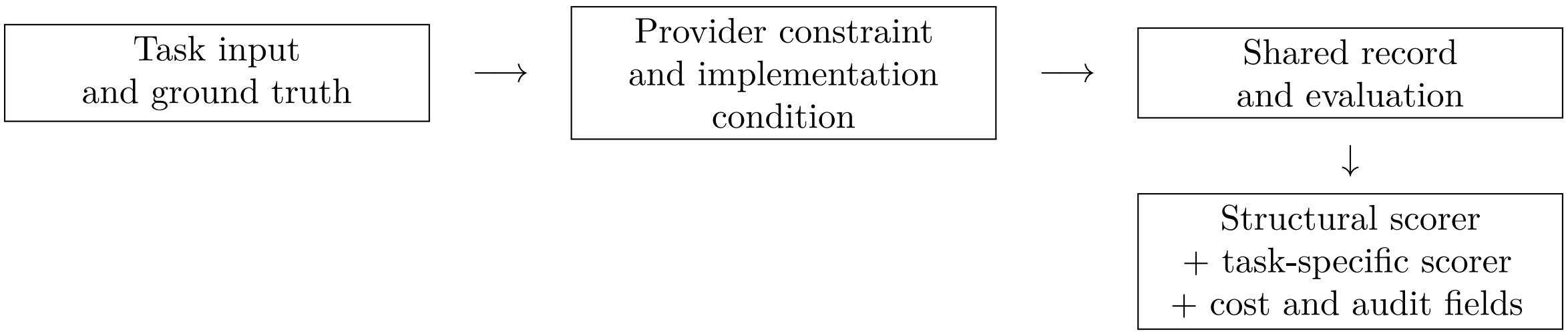

\centering
\small
\setlength{\tabcolsep}{5pt}
\begin{tabularx}{\textwidth}{>{\centering\arraybackslash}X c >{\centering\arraybackslash}X c >{\centering\arraybackslash}X}
\fbox{\parbox{0.25\textwidth}{\centering Task input\\and ground truth}}
& $\longrightarrow$ &
\fbox{\parbox{0.25\textwidth}{\centering Provider constraint\\and implementation condition}}
& $\longrightarrow$ &
\fbox{\parbox{0.25\textwidth}{\centering Shared record\\and evaluation}}\\[0.4em]
&&&& $\downarrow$ \\[0.3em]
&&&& \fbox{\parbox{0.25\textwidth}{\centering Structural scorer\\
$+$ task-specific scorer\\
$+$ cost and audit fields}}
\end{tabularx}
\caption{Layered evaluation path (conceptual figure). Provider constraints, framework behavior, application validation, and task scoring are distinct mechanisms. Feature-using ablations exercise validators, typed dependencies, stateful orchestration, and optimization beyond the lowest-common-denominator structured-output path.}
\label{fig:evaluation-layers}
\end{figure}

Figure~\ref{fig:evaluation-layers} locates the layers in the benchmark. Structural and task-specific scoring remain separate after every implementation returns. This shared evaluation boundary is necessary because a framework-native ``success'' may mean only that an API call completed.

The comparison conditions occupy different parts of this path. LangGraph emphasizes explicit state and recoverable orchestration \citep{langgraph2026}; DSPy emphasizes metric-driven optimization \citep{khattab2023dspy,dspyDocs2026}; CrewAI emphasizes role and task decomposition \citep{crewai2026}; Agno provides a structured-output interface \citep{agno2026}; and the plain SDK reference exposes provider behavior with minimal framework abstraction. A fair protocol must therefore distinguish a matched lowest-common-denominator condition from feature-using ablations. The former supports controlled execution comparisons; the latter are needed to attribute effects to validators, optimization, graph state, or typed tools.

\section{Related Work}
\label{sec:related_work}

Structured-output research increasingly distinguishes representation from value correctness. JSONSchemaBench, StructEval, SOB, LLMStructBench, and ExtractBench measure schema coverage, parseability, and leaf- or field-level accuracy as separable outcomes \citep{geng2025jsonschemabench,yang2025structeval,singh2026sob,tenckhoff2026llmstructbench,ferguson2026extractbench}. Recent measurements sharpen the same point operationally: hard schema constraints can raise validity while increasing wrong-but-schema-valid answers, so production reports should track validity, answer accuracy, executable accuracy, and wrong-valid-schema rate separately \citep{ray2026constrainttax}. Instructor, Guardrails, TypeChat, LMQL, guided decoding, SchemaRL, and SLOT place enforcement at different points---provider decoding, grammar constraints, post-hoc structuring, reinforcement against validators, or application repair \citep{liu2024instructor,guardrails2026,typechat2026,beurerkellner2023lmql,willard2023guided,lu2025schemarl,li2025slot}. Empirically, verify-and-retry against an external validator can cover more real-world JSON Schemas than brittle constrained-decoding stacks alone \citep{lee2025goodenough}, and software-engineering studies of structured-output control likewise treat the binding layer as an engineering choice rather than a solved default \citep{song2026structuredoutputcontrol}. This work motivates our separation of final schema validity from task-specific scoring and our attention to where enforcement runs.

Agent-evaluation research broadens the unit of analysis beyond the final answer. AgentBench, AgentBoard, BFCL, and TauBench show that tool calls, state transitions, cost, and implementation details affect observed reliability \citep{liu2024agentbench,ma2024agentboard,patil2025bfcl,yao2025taubench}. Path- and validator-centric evaluations go further: CORE scores full tool-use trajectories rather than terminal state alone \citep{michelakis2025core}, and deterministic structured validation of tool traces can overturn rankings based on string heuristics or LLM judges while exposing schema-valid but wrong arguments and fabricated tool use \citep{gurram2026toolvalidator}. Surveys similarly caution that framework abstractions, orchestration, and evaluation infrastructure can be part of the measured system rather than neutral plumbing \citep{mohammadi2025agentsurvey,yehudai2025agentsurvey,kapoor2024agentsmatter}. We therefore record implementation condition, task, sample, retries, latency, token counts when exposed, parsed output, task scores, and errors in a common row, and we score executed workflow evidence rather than proposal-only success.

\begin{table}[t]
\centering
\small
\begin{tabularx}{\textwidth}{P{0.22\textwidth}P{0.24\textwidth}Y}
\toprule
Benchmark family & Typical unit of analysis & Gap relative to this protocol \\
\midrule
Constrained decoding / JSON Schema benches (e.g., JSONSchemaBench, StructEval, SOB) & Schema coverage, parseability, constrained-decoding engines & Rarely compares application-boundary implementations on matched business tasks with shared task scorers. \\
Validity--correctness studies (e.g., Constraint Tax) & Paired validity vs.\ answer/executable accuracy & Strong on metric separation; not a multi-framework business-workflow harness. \\
Agent tool/state benches (e.g., AgentBench, CORE, tool-trace validators) & Tools, trajectories, path or state success & Often omit a shared schema-validity surface and an SDK reference that isolates provider enforcement. \\
Business-task evals (SQL, extraction, factuality) & Task correctness on domain data & Often report correctness without comparing the engineering boundary that accepted the output. \\
This protocol & Implementation condition $\times$ business task under shared schemas, scorers, and provenance & Connects structural validity, task scoring, cost, and mechanism ablations under a frozen confirmatory plan. \\
\bottomrule
\end{tabularx}
\caption{Related-work contrast. Structured-output benches emphasize decoding engines; validity--correctness studies separate metrics; agent benches emphasize tools and paths; this protocol compares application implementations on matched workflow contracts.}
\label{tab:related-work-comparison}
\end{table}

The task families draw on established data and business benchmarks. SQL generation follows execution-based evaluation traditions from Spider, BIRD, and Spider 2.0 \citep{yu2018spider,li2023bird,lei2025spider2,kim2024flex}. Document extraction follows form and invoice benchmarks such as FUNSD, SROIE, and DocILE \citep{jaume2019funsd,huang2019sroie,simsa2023docile}. Compliance question answering and report summarization inherit grounding and hallucination concerns from retrieval-augmented generation and factuality research \citep{lewis2020rag,ji2023hallucination}.

Finally, benchmark cards and evaluation cards motivate explicit provenance and scope metadata \citep{sokol2025benchmarkcards,ghosh2026evaluationcards}. Security and AI-management guidance likewise emphasizes traceable controls, monitoring, and safe handling of model output rather than one-time approval \citep{nistGenAIProfile2024,owaspLLMTop102025,iso42001_2023}.

The gap lies between these literatures. Structured-output benchmarks usually isolate schema adherence or constrained-decoding engines, while agent benchmarks emphasize tools and trajectories; business-task evaluations often report correctness without comparing the engineering boundary that accepted the output. This study connects schema enforcement, task-specific scoring, and implementation cost under matched workflow inputs across six Python conditions, with a confirmatory matrix under a frozen analysis plan and labeled ablations for mechanism claims.

\section{Research Questions and Hypotheses}
\label{sec:research_questions}

The confirmatory study is organized around four questions and three pre-specified hypotheses about schema-validated business workflows. The hypotheses are stated at the \emph{system-configuration} level: they concern matched integration paths under a shared model, harness, and scorers, not causal claims about marketed library features unless a labeled ablation activates that feature.

\paragraph{RQ1: Execution coverage.}
Can every configured implementation--task path complete live API execution and emit a row that can be scored and aggregated by the shared reporting pipeline?

\paragraph{RQ2: Measurement divergence.}
Under schema-validated execution, can an output recorded as schema valid still receive a lower task-specific score, and which candidate discrepancies does the deterministic audit procedure surface?

\paragraph{RQ3: Comparative effects.}
After hardening the task instances, controls, and attempt-level instrumentation, how do implementation conditions differ on structural validity and task-specific correctness for matched workflow inputs?

\paragraph{RQ4: Mechanisms and cost.}
Which observed differences can be attributed to provider enforcement, application validation, retry feedback, typed dependencies, optimization, or orchestration, and what latency, token, implementation, and maintenance costs accompany them?

\paragraph{H1 (separability).}
Structural validity and task-specific correctness are separable outcomes: non-trivial rates of schema-valid rows will still fall below the automated quality threshold, especially on extraction and compliance tasks where fluent objects can omit entities or unsupported claims.

\paragraph{H2 (binding-path heterogeneity).}
Under a shared model and lowest-common-denominator structured-output contract, implementation conditions will not be interchangeable. Typed or structured-binding paths (PydanticAI, LangGraph, DSPy in this package state) will outperform a plain SDK baseline and weaker adapter bindings on schema validity and automated quality.

\paragraph{H3 (condition$\times$task interaction).}
Performance gaps are not uniform additive shifts. Condition and task will interact: some tasks (e.g., summarization) compress differences, while others (e.g., workflow routing) amplify adapter failures even when the same condition succeeds elsewhere.

RQ1 and RQ2 are answered by execution evidence and diagnostic contingency tables that report effective design counts alongside row counts. RQ3 is answered by the eligible confirmatory matrix under the frozen analysis plan in Table~\ref{tab:power-summary}, with H1--H3 evaluated against that matrix. RQ4 requires labeled feature-using ablations and must not be inferred from LCD structured-output conditions alone. In particular, workflow claims require executed tool/state evidence, and DSPy optimization claims require the held-out optimized ablation rather than the untuned primary condition.

Table~\ref{tab:rq-scope} makes the evidence boundary explicit.

\begin{table}[t]
\centering
\small
\begin{tabularx}{\textwidth}{P{0.12\textwidth}P{0.25\textwidth}Y}
\toprule
Question & Current evidence & Admissible claim \\
\midrule
RQ1 & Hardened runners, provenance, and live execution paths across 30 implementation--task cells & Configured paths complete and enter the reporting pipeline under pinned model/package settings. \\
RQ2 / H1 & Structural flags, deterministic task scores, effective design counts & Schema validity can coexist with lower automated task scores; flags are candidates until human review. \\
RQ3 / H2--H3 & Eligible confirmatory matrix under frozen power/equivalence settings & Instance-clustered paired contrasts vs plain SDK; condition$\times$task residuals describe interactions. \\
RQ4 & Labeled ablation/stress rows and mechanism tables & Causal attribution only for conditions that activate the claimed mechanism. \\
\bottomrule
\end{tabularx}
\caption{Research-question and hypothesis evidence boundary for the confirmatory study.}
\label{tab:rq-scope}
\end{table}

\section{Benchmark Design}
\label{sec:benchmark}

\subsection{Design Overview}

The benchmark uses matched inputs: each implementation condition receives the same task record, output schema, model family, and nominal retry limit. Framework-specific code binds these objects using the library's native interface. The unit for comparative analysis is a distinct task instance, with repeated provider calls nested within that instance. The row field \texttt{seed} is treated as a replicate identifier unless a provider seed is actually set and verified.

All six confirmatory conditions use explicit structured outputs. The primary matrix therefore compares integration paths under provider- or library-mediated schema enforcement; it does not compare typed output with unconstrained text. Mechanism claims require separate labeled conditions that change provider enforcement, application validation, validators, tools, graph state, or optimization while preserving the common scoring boundary.

\subsection{Tasks}

Table~\ref{tab:tasks} summarizes the five task families. The name \emph{workflow decision} is used because the LCD condition predicts a classification, tool name, and review flag; the confirmatory scorer additionally executes shared tool stubs and checks state-path evidence when present.

\begin{table}[t]
\centering
\small
\begin{tabularx}{\textwidth}{P{0.19\textwidth}Y Y}
\toprule
Task & Input & Required output \\
\midrule
Structured extraction & Synthetic invoices and complaints & Typed fields, optional values, confidence, and error state \\
SQL generation & Question and SQLite schema & Query, explanation, referenced tables and columns, or refusal \\
Report summarization & Synthetic tabular and narrative report & Headline, revenue summary, risks, recommendations, and outlook \\
Compliance QA & Synthetic policy and question & Answer, quoted citations, confidence, unsupported-claim list, and review flag \\
Workflow decision & Synthetic support request & Classification, proposed tool, extracted entities, recommendation, and review flag \\
\bottomrule
\end{tabularx}
\caption{Task families in the current artifact.}
\label{tab:tasks}
\end{table}

\subsection{Implementation Conditions}

Table~\ref{tab:conditions} describes what the confirmatory LCD matrix exercises. This prevents framework capability from being inferred from features that were not active in the primary run.

\begin{table}[t]
\centering
\small
\begin{tabularx}{\textwidth}{P{0.16\textwidth}P{0.32\textwidth}Y}
\toprule
Condition & Confirmatory LCD mechanism & Interpretation limit \\
\midrule
PydanticAI & \texttt{Agent} with a Pydantic \texttt{output\_type} & Validators/typed deps reserved for labeled ablations. \\
Plain SDK & Provider JSON Schema response format followed by Pydantic parsing & Framework-free reference for provider enforcement plus local parsing. \\
LangGraph & Structured model call wrapped in a small retry graph & Does not test a stateful business process or human interrupt. \\
DSPy & Untuned typed signature and \texttt{Predict} with application coercion & Does not test metric-driven optimization. \\
CrewAI & One agent and one task with \texttt{output\_pydantic} & Does not test delegation or multi-agent coordination. \\
Agno & One agent with a Pydantic output schema & Does not test tools, teams, or workflow state. \\
\bottomrule
\end{tabularx}
\caption{Mechanisms exercised by the confirmatory lowest-common-denominator matrix.}
\label{tab:conditions}
\end{table}

\subsection{Structural and Task-Specific Outcomes}

For implementation condition $f$, task $t$, and rows $D_{f,t}$, final schema validity is
\begin{equation}
\operatorname{FinalValid}(f,t)=
\frac{1}{|D_{f,t}|}\sum_{i\in D_{f,t}}\mathbb{1}[v_i=1],
\end{equation}
where $v_i$ is the harness-boundary schema-valid flag after independent revalidation. Zero-retry success additionally requires the observed provider-call count to equal one (or recorded retries $r_i=0$ when calls are unavailable):
\begin{equation}
\operatorname{ZRS}(f,t)=
\frac{1}{|D_{f,t}|}\sum_{i\in D_{f,t}}\mathbb{1}[v_i=1\land c_i=1],
\end{equation}
where $c_i$ is observed \texttt{provider\_calls}.

Task-specific scores are computed after the runner returns. Table~\ref{tab:task-metrics} lists the confirmatory primary scores.

\begin{table}[t]
\centering
\small
\begin{tabularx}{\textwidth}{P{0.18\textwidth}P{0.33\textwidth}Y}
\toprule
Task & Confirmatory primary score & Hardening in the artifact \\
\midrule
Extraction & Macro-average of normalized field matches & Distinct layouts/templates with \texttt{template\_id}; label-derivability tests \\
SQL & Result-set equality under read-only checks & Separate fresh DBs for predicted vs gold; reject non-SELECT writes \\
Summarization & Source-fact coverage with unsupported-content penalty & Forbidden-term penalty in quality score; blinded review packet \\
Compliance QA & Answer/citation support with unsupported-claim penalty & Claim-grounding scores beyond empty-list exact match \\
Workflow & Decision fields plus tool execution and state-path match & Shared tool stubs and expected state paths scored on every row \\
\bottomrule
\end{tabularx}
\caption{Task-specific confirmatory scorers and hardening implemented in the executable artifact.}
\label{tab:task-metrics}
\end{table}

Efficiency fields include elapsed time and provider token counts when exposed. Engineering fields include runner file count, nonblank lines, optional dependencies, and framework-specific markers. They describe the artifact surface, not developer productivity or maintainability.

\subsection{Normalized Record and Scoring Boundary}

Each call is serialized to a common row containing implementation condition, task, sample identifier, replicate label, harness-boundary schema validity, observed provider calls, latency, token counts, parsed output, a raw-output field when exposed, task sub-scores, a primary diagnostic score, an error string, and an optional trace-event list. The harness distinguishes runner-reported acceptance from independent revalidation against the committed task schema and attaches provenance hashes for configuration and scorers.

\subsection{Fairness Rules}

The protocol fixes shared task content, model settings, a maximum provider-call budget, schemas, and scoring code within each comparison. It randomizes or counterbalances condition order within matched input blocks and archives every deviation required by a library. DSPy optimization uses a development split and reports tuning calls separately; stateful and tool-using conditions report the additional actions they perform rather than being forced into a lowest-common-denominator label. Feature-using ablations are labeled separately from the primary matrix.

\section{Experimental Setup and Analysis Plan}
\label{sec:experimental_setup}

\subsection{Confirmatory Configuration}

The confirmatory run uses \texttt{openai:gpt-4o-mini}, six implementation conditions, five task families, and three nested replicates on distinct label-derivable instances ($100$ extraction and $50$ per other task):
\begin{equation}
6\ \text{conditions}\times 3\ \text{replicates}\times
(100+50+50+50+50)\ \text{task rows}=5{,}400.
\end{equation}
Effective sample size for comparative estimands is the number of distinct instances, not the nominal row total. The term \texttt{seed} in the row schema is retained for backward compatibility and interpreted as a replicate identifier unless a provider seed is actually set and verified.

Eligibility required: (1) a common maximum number of observed provider calls; (2) attempt-level retry and validation events; (3) uniform propagation of supported model settings; (4) independently revalidated outputs; (5) randomized or counterbalanced condition order within input blocks; (6) read-only SQL execution on separate fresh databases; and (7) stronger grounding and hallucination measures. Separately labeled feature-using ablations (validators, typed dependencies, retry budgets, orchestration, DSPy optimization) answer RQ4 and are not required to report the LCD confirmatory matrix. The eligible run \texttt{confirmatory-5400-v1} satisfies the primary-matrix gate under the frozen analysis plan.

Automated task scores are computed from stored parsed objects under deterministic scorers. Rows below the $0.99$ quality threshold are reported as \emph{metric-flagged candidates}, not confirmed semantic errors, until blinded human adjudication is complete.

\subsection{Estimands and Hypotheses}

The primary unit is a distinct task instance. Repeated calls are averaged within instance for continuous outcomes and retained as nested Bernoulli observations for structural outcomes. Each condition is compared with the framework-free SDK reference on the same instances; PydanticAI--Agno is the pre-specified secondary contrast for schema-oriented interfaces. Other pairwise comparisons are exploratory.

The primary structural estimands are the paired difference in independently verified first-call validity and final validity after a common call budget. The task estimand is a paired difference in the pre-specified task score after the metric hardening described in Table~\ref{tab:task-metrics}. Cost estimands include provider calls, tokens with a coverage indicator, and elapsed time after blocking or adjustment for randomized execution order. Failed calls remain in the intention-to-run denominator.

H1 is evaluated with schema-valid $\times$ quality-flag contingencies by task. H2 is evaluated with condition means and paired contrasts versus the SDK. H3 is evaluated with condition$\times$task quality residuals after removing additive main effects. Confidence intervals resample distinct instances. Within each task, comparisons against the SDK reference use a Holm adjustment; unadjusted intervals and secondary diagnostics remain in the artifact. Equivalence margins and simulation-based power were frozen in \texttt{benchmark/config/analysis\_plan.json} before launch. Table~\ref{tab:power-summary} summarizes approximate MDEs and equivalence power.

\begin{table}[t]
\centering
\small
\begin{tabularx}{\textwidth}{lrrrY}
\toprule
Task & Instances & Approx.\ MDE (quality) & Equiv.\ power @ 0 & Notes \\
\midrule
Extraction & 100 & 0.024 & 0.999 & 3 nested replicates \\
SQL generation & 50 & 0.034 & 0.986 & Distinct DB per instance \\
Summarization & 50 & 0.034 & 0.986 & Unsupported-content penalty \\
Compliance QA & 50 & 0.034 & 0.986 & Claim grounding \\
Workflow & 50 & 0.034 & 0.986 & Executed tool/state scoring \\
\bottomrule
\end{tabularx}
\caption{Frozen instance-clustered power diagnostics from \texttt{config/analysis\_plan.json}. MDEs are approximate paired mean differences for quality\_score at 80\% power (sd~$=0.15$, $\alpha=0.05$). Equivalence margins are $\pm 0.05$. These values replace nominal 5{,}400-row arithmetic as the confirmatory sample-size story.}
\label{tab:power-summary}
\end{table}

\subsection{Engineering Outcomes}

Static runner counts are reported in Appendix~\ref{sec:implementation_patterns} rather than in the primary results, because adapter LOC and API-marker tallies describe artifact surface area only. Maintainability claims require a blinded schema-change exercise with recorded completion time, defects, and reviewer ratings. Traceability claims require populated attempt and process events; empty \texttt{trace\_events} fields are not evidence of trace completeness.

\section{Failure Classification}
\label{sec:failure_taxonomy}

The protocol classifies evidence by the first observable layer at which a run departs from the intended contract:

\begin{description}
\item[Transport or compatibility.] Authentication, rate limits, timeouts, unsupported provider schemas, missing dependencies, or incompatible package APIs prevent an evaluable response. These are deployment failures, but they are not task-correctness errors.
\item[Parse or schema.] A response cannot be decoded, fails the declared runtime model, or exhausts the common repair budget. Confirmatory records must distinguish provider rejection, framework-native validation, and independent benchmark revalidation.
\item[Task discrepancy.] A schema-valid object differs from deterministic ground truth or receives a lower task-specific score. A threshold rule can select \emph{candidate} discrepancies for audit; the label \emph{confirmed error} is reserved for deterministic checks with valid ground truth or completed human adjudication.
\item[Process or state.] A system chooses an unauthorized or incorrect tool, skips a review gate, loses state, or reports success without a supporting tool outcome. This category requires executed actions and process traces; the current single-call workflow-decision task can only flag proposed-decision mismatches.
\item[Silent degradation.] A run is accepted by the application boundary but lacks enough evidence to support its disposition. Examples include a quote that appears in the source but does not entail the answer, an unsupported summary claim, or a valid object whose raw request/attempt evidence is missing.
\end{description}

This taxonomy is diagnostic rather than mutually exclusive. One row may contain both a task discrepancy and a process flag. Counts are therefore reported per category with overlap made explicit, and deterministic selection rules are preserved alongside the selected examples.

\section{Results}
\label{sec:results}

\subsection{Confirmatory Primary Matrix (RQ1--RQ3)}

The eligible confirmatory run \texttt{confirmatory-5400-v1} completed 5{,}400 unique execution rows under the frozen analysis plan: six implementation conditions, five tasks, distinct instance counts of $100+50+50+50+50$, and three nested replicates, with \texttt{pilot=false}. Every row carries a \texttt{template\_id}, harness-boundary schema revalidation flags, observed \texttt{provider\_calls}, and scorer/provenance hashes. RQ1 is therefore answered affirmatively for the configured matrix: all 30 implementation--task cells entered the shared reporting pipeline.

Overall schema validity was 0.684 and mean automated quality was 0.638. Tables~\ref{tab:conf-schema-matrix} and~\ref{tab:conf-quality-matrix} show that these averages conceal a sharp two-regime pattern. PydanticAI and LangGraph achieved ceiling schema validity (100\%) with mean quality $0.935$--$0.940$; DSPy was nearly identical (99.8\% schema validity; quality $0.911$). Plain SDK and Agno clustered near 39\% schema validity and quality $\approx 0.36$, while CrewAI was lowest overall (32.8\% schema validity; quality $0.308$), including a complete structural collapse on the workflow task (0\% schema-valid).

\begin{table}[t]
\centering
\resizebox{\textwidth}{!}{%
\small
\begin{tabular}{lrrrrrr}
\toprule
Condition & Extract & SQL & Summary & Compliance & Workflow & Overall \\
\midrule
PydanticAI & 100.0 & 100.0 & 100.0 & 100.0 & 100.0 & 100.0 \\
LangGraph & 100.0 & 100.0 & 100.0 & 100.0 & 100.0 & 100.0 \\
DSPy & 99.3 & 100.0 & 100.0 & 100.0 & 100.0 & 99.8 \\
Plain SDK & 33.3 & 33.3 & 64.0 & 35.3 & 33.3 & 38.8 \\
Agno & 33.3 & 33.3 & 63.3 & 35.3 & 34.0 & 38.8 \\
CrewAI & 32.0 & 33.3 & 64.0 & 35.3 & 0.0 & 32.8 \\
\midrule
Task mean & 66.3 & 66.7 & 81.9 & 67.7 & 61.2 & 68.4 \\
\bottomrule
\end{tabular}}
\caption{Confirmatory schema-validity rates (\%) by implementation condition and task. Overall is row-weighted across the 5{,}400-row matrix.}
\label{tab:conf-schema-matrix}
\end{table}

\begin{table}[t]
\centering
\resizebox{\textwidth}{!}{%
\small
\begin{tabular}{lrrrrrr}
\toprule
Condition & Extract & SQL & Summary & Compliance & Workflow & Overall \\
\midrule
PydanticAI & 0.921 & 0.873 & 0.992 & 0.973 & 0.928 & 0.935 \\
LangGraph & 0.922 & 0.880 & 1.000 & 0.976 & 0.941 & 0.940 \\
DSPy & 0.890 & 0.760 & 1.000 & 0.981 & 0.944 & 0.911 \\
Plain SDK & 0.307 & 0.293 & 0.640 & 0.345 & 0.308 & 0.367 \\
Agno & 0.307 & 0.287 & 0.631 & 0.345 & 0.313 & 0.365 \\
CrewAI & 0.292 & 0.280 & 0.640 & 0.345 & 0.000 & 0.308 \\
\midrule
Task mean & 0.607 & 0.562 & 0.817 & 0.661 & 0.572 & 0.638 \\
\bottomrule
\end{tabular}}
\caption{Confirmatory mean automated quality by implementation condition and task. Quality metrics are task-specific (field accuracy, execution, grounding, etc.) under the shared scorers.}
\label{tab:conf-quality-matrix}
\end{table}

These contrasts support H2 at the system-configuration level: under a shared model and LCD structured-output contract, binding paths are not interchangeable. They are \emph{not} causal attributions to named library features; feature-using ablations are required for mechanism claims (RQ4).

\subsection{Hypothesis Checks: Separability and Interactions}

Table~\ref{tab:rq2-contingency} evaluates H1. Across tasks, large numbers of schema-valid rows still fall below the $0.99$ automated quality threshold---especially extraction (642 flags among 1{,}194 schema-valid rows) and compliance QA (370 among 609). Summarization is the exception: high schema validity coincides with few flags and the highest mean quality ($0.817$). Schema acceptance therefore cannot be read as task success.

\begin{table}[t]
\centering
\small
\begin{tabularx}{\textwidth}{lrrrrY}
\toprule
Task & Rows & Designs & Schema-valid & Flagged ($q<0.99$) & Mean $q$ \\
\midrule
compliance\_qa & 900 & 50 & 609 & 370 & 0.661 \\
extraction & 1800 & 100 & 1194 & 642 & 0.606 \\
multi\_step & 900 & 50 & 551 & 175 & 0.572 \\
sql\_generation & 900 & 50 & 600 & 94 & 0.562 \\
summarization & 900 & 50 & 737 & 9 & 0.817 \\
\bottomrule
\end{tabularx}
\caption{RQ2 contingency: schema-valid rows versus metric-flagged task discrepancies, with effective design counts (distinct sample\_id).}
\label{tab:rq2-contingency}
\end{table}

Table~\ref{tab:conf-quality-interaction} evaluates H3 using additive residuals $q_{ft}-\bar q_f-\bar q_t+\bar q$. Positive cells outperform what condition and task means alone would predict; negative cells underperform. Two patterns stand out. First, summarization shows large positive residuals for the weaker binders (Plain SDK, Agno, CrewAI) and negative residuals for the strong binders: when the task is easier, the binding gap compresses. Second, CrewAI's workflow residual ($-0.243$) and DSPy's SQL residual ($-0.076$) show that failures concentrate in particular condition--task cells rather than as uniform level shifts. Interaction, not only main effects, is part of the confirmatory story.

\begin{table}[t]
\centering
\resizebox{\textwidth}{!}{%
\small
\begin{tabular}{lrrrrr}
\toprule
Condition & Extract & SQL & Summary & Compliance & Workflow \\
\midrule
PydanticAI & +0.017 & +0.014 & -0.122 & +0.015 & +0.058 \\
LangGraph & +0.013 & +0.015 & -0.120 & +0.013 & +0.066 \\
DSPy & +0.011 & -0.076 & -0.091 & +0.047 & +0.098 \\
Plain SDK & -0.028 & +0.002 & +0.094 & -0.045 & +0.006 \\
Agno & -0.027 & -0.003 & +0.087 & -0.043 & +0.014 \\
CrewAI & +0.015 & +0.047 & +0.152 & +0.014 & -0.243 \\
\bottomrule
\end{tabular}}
\caption{Condition$\times$task interaction residuals for mean quality after removing additive condition and task effects ($q_{ft}-\bar q_f-\bar q_t+\bar q$). Positive cells outperform the additive baseline; negative cells underperform.}
\label{tab:conf-quality-interaction}
\end{table}

\subsection{Paired Contrasts versus the SDK Reference}

Paired instance-clustered contrasts versus the plain SDK (900 pairs, 300 clusters) are archived under \texttt{inferential/}. Pre-specified quality contrasts reinforce the two-regime pattern: PydanticAI minus plain SDK had mean difference $0.568$ with 95\% bootstrap CI $[0.550,0.585]$ (550/17/333 a-better/b-better/ties); LangGraph $0.573$ $[0.556,0.590]$; DSPy $0.544$ $[0.522,0.566]$. Agno was statistically compatible with the SDK ($-0.002$ $[-0.006,0.002]$), while CrewAI was worse ($-0.059$ $[-0.073,-0.045]$). Analogous JSON summaries are retained with the result bundle.

\subsection{Ablations for Mechanism Evidence (RQ4)}

Labeled ablations test whether the confirmatory gap is explained by retries, domain validators, or typed dependencies rather than by the LCD binding path alone.

Table~\ref{tab:ablation-zero-retry} reports zero-retry versus final-after-budget conditions across all six implementations. Strong binders already succeed on the first call; weaker binders consume additional provider calls under the retry budget without matching the confirmatory-scale quality of PydanticAI/LangGraph/DSPy. Retries are therefore a cost center for fragile adapters, not a substitute for a reliable structured-output binding.

\begin{table}[t]
\centering
\resizebox{\textwidth}{!}{%
\small
\begin{tabular}{llrrrrrr}
\toprule
Condition & Metric & PydanticAI & LangGraph & DSPy & Plain SDK & Agno & CrewAI \\
\midrule
Zero-retry & schema & 100.0 & 100.0 & 80.0 & 100.0 & 80.0 & 60.0 \\
 & quality & 0.925 & 0.934 & 0.741 & 0.926 & 0.730 & 0.555 \\
 & provider\_calls & 1.00 & 1.00 & 1.00 & 1.00 & 1.00 & 1.00 \\
\midrule
After retry budget & schema & 100.0 & 100.0 & 80.0 & 100.0 & 80.0 & 60.0 \\
 & quality & 0.923 & 0.934 & 0.741 & 0.926 & 0.739 & 0.555 \\
 & provider\_calls & 1.00 & 1.00 & 1.60 & 1.00 & 1.60 & 2.20 \\
\bottomrule
\end{tabular}}
\caption{Updated zero-retry versus final-after-budget ablation (pilot matrix: 5 samples $\times$ 5 tasks, one replicate). Retry budget increases mean provider calls for weaker binders without closing the confirmatory-scale gap to strong binders.}
\label{tab:ablation-zero-retry}
\end{table}

Tables~\ref{tab:ablation-validators} and~\ref{tab:ablation-typed} isolate feature flags on PydanticAI. Enabling domain output validators leaves schema validity at ceiling and yields small quality lifts on compliance and workflow (Table~\ref{tab:ablation-validators}). Typed dependencies are visible in traces and produce a modest overall quality increase relative to prompt-only context (Table~\ref{tab:ablation-typed}). These mechanism increments are an order of magnitude smaller than the confirmatory gap between strong and weak binders, which supports treating H2 as primarily a binding-path finding while reserving larger RQ4 claims for broader feature-using and secondary-model campaigns.

\begin{table}[t]
\centering
\resizebox{\textwidth}{!}{%
\small
\begin{tabular}{llrrr}
\toprule
Condition & Task & Schema\% & Quality & Calls \\
\midrule
Validators off & SQL & 100.0 & 0.867 & 1.00 \\
 & Compliance & 100.0 & 0.966 & 1.00 \\
 & Workflow & 100.0 & 0.920 & 1.00 \\
\midrule
Validators on & SQL & 100.0 & 0.867 & 1.00 \\
 & Compliance & 100.0 & 0.973 & 1.00 \\
 & Workflow & 100.0 & 0.933 & 1.00 \\
\bottomrule
\end{tabular}}
\caption{PydanticAI domain-validator ablation (15 instances $\times$ 3 replicates per cell). Enabling output validators yields small quality gains on compliance and workflow under an already ceiling schema-validity regime.}
\label{tab:ablation-validators}
\end{table}

\begin{table}[t]
\centering
\resizebox{\textwidth}{!}{%
\small
\begin{tabular}{llrrrr}
\toprule
Condition & Task & Schema\% & Quality & Calls & Typed-deps trace \\
\midrule
Prompt-only context & Compliance & 100.0 & 0.973 & 1.00 & 0\% \\
 & Workflow & 100.0 & 0.911 & 1.00 & 0\% \\
\midrule
Typed dependencies & Compliance & 100.0 & 0.974 & 1.00 & 100\% \\
 & Workflow & 100.0 & 0.933 & 1.00 & 100\% \\
\bottomrule
\end{tabular}}
\caption{PydanticAI typed-dependency ablation (15 instances $\times$ 3 replicates). Typed deps are observable in traces; quality differences remain small relative to the confirmatory LCD gap versus weaker binders.}
\label{tab:ablation-typed}
\end{table}

\section{Discussion}
\label{sec:discussion}

\subsection{From Hypotheses to Discriminating Contrasts}

The confirmatory matrix was designed to test separability, binding-path heterogeneity, and interaction---not to crown a framework winner. After hardening (distinct instances, observed call budgets, boundary revalidation, and executed workflow scoring), the matrix produced discriminating structural and quality contrasts under a shared model and scoring boundary, consistent with H1--H3. The same run also validates the protocol machinery itself: instance-level analysis, nested replicates, frozen power and equivalence settings, harness-boundary validation, and attempt-level provenance. Nominal 5{,}400-row arithmetic is tied to that frozen plan rather than treated as effective $N$ by itself.

The pattern that PydanticAI, LangGraph, and DSPy retained higher recorded schema validity than the plain SDK, Agno, and CrewAI conditions in this package state is a system-level finding supporting H2. It should not be read as a claim that any library's marketed features caused the difference: the primary matrix remains an LCD structured-output comparison, and CrewAI/Agno adapters remain sensitive to dependency and API surface changes. Feature-using ablations (validators, typed dependencies, stateful graphs, optimized DSPy) are the correct attribution step for RQ4.

\subsection{What the Technology Stack Is Doing}

The six conditions are best read as different ways of binding the same business contract to a model call, not as interchangeable ``agent frameworks.''

Modern APIs can constrain generation to a JSON Schema. That layer is necessary and powerful, but it is still a generation constraint: it increases the chance of a parseable object; it does not prove that invoice totals, SQL predicates, policy citations, or routing decisions are correct. The plain SDK reference exists to measure this baseline without application-framework mediation.

Pydantic models, and agent wrappers built around them (here, PydanticAI), turn the contract into code the rest of the system can import, validate, and reject. In practice this changes the failure mode: instead of silently accepting a loosely shaped dict, the application can fail closed, retry with validation feedback, or route to review---consistent with evidence that external validate-and-retry loops can be more robust than relying on constrained decoding alone for complex schemas \citep{lee2025goodenough}. Near-ceiling schema validity in the confirmatory matrix for PydanticAI is therefore evidence that the typed boundary held under our harness---not evidence that field values were true.

LangGraph-style graphs make control flow explicit: retry loops, conditional branches, tool nodes, and later human interrupts become inspectable state transitions. In the LCD confirmatory matrix the LangGraph adapter is still primarily a structured-call-plus-retry wrapper, so the high structural scores speak to reliable structured binding under that wrapper rather than to long-running stateful processes. Graphs buy auditability and recoverable control when the workflow itself is multi-step; they are overkill if the only need is a single validated object. DSPy, by contrast, treats prompts and intermediate structure as tunable programs evaluated against metrics. Untuned DSPy in the primary matrix behaves as another structured-output integration path; optimized holdouts are required before claiming that compilation or few-shot bootstrapping improved reliability.

CrewAI and Agno emphasize role/task composition or a compact agent surface. Their lower confirmatory schema-validity and quality under this package state illustrate a different risk: framework convenience layers can change how schemas are passed, how tools are exposed, or how retries are counted. The lesson is not that those libraries cannot be made reliable, but that ``supports structured output'' in documentation is not the same as ``survives a shared harness boundary.'' Adapter and dependency pin choices are part of the reliability story.

A central design choice in this artifact is independent revalidation at the benchmark boundary. Framework-reported success, provider acceptance, and application-schema acceptance can diverge. Path-aware agent evaluation makes the same point for tool sequences: terminal success can hide unsafe or inefficient intermediate actions \citep{michelakis2025core}. Teams shipping business workflows need the same discipline: log what the provider returned, what the local schema accepted, and what the task scorer or reviewer decided, as three separate events.

\subsection{What Lower Task Scores Still Show}

Even among schema-valid rows, Table~\ref{tab:rq2-contingency} shows metric-flagged discrepancies, especially in extraction and compliance QA. Runtime acceptance and automated task scoring continue to disagree. Blinded human adjudication of the confirmatory audit packet is required before interpreting flags as confirmed semantic error rates. The same divergence is now a recurring theme in structured-output measurement: Constraint Tax shows hard schemas can inflate wrong-valid-schema rates even as validity approaches 100\% \citep{ray2026constrainttax}, and tool-agent audits show schema-valid wrong arguments can propagate into downstream failures that final-answer heuristics miss \citep{gurram2026toolvalidator}.

Technically, this split is expected. Extraction and compliance are where fluent, schema-shaped objects can still omit entities, invent unsupported claims, or miss grounding spans. Summarization's much lower flag rate in the same matrix suggests that some task contracts are easier for current scorers---or easier for models under schema constraints---than others. Deployment teams should therefore budget evaluation effort by task risk, not by a single global ``structured output works'' dashboard, and should report validity and task correctness as separate operating metrics.

\subsection{Implications for Engineering Practice}

For engineering practice, the confirmatory picture supports a layered stack rather than a framework beauty contest. Start from an explicit schema and an SDK (or minimal) structured-output path so provider behavior is visible; add application validation when rejected or repaired objects are cheaper than silent corruption; add orchestration when the business process itself needs state, tools, approvals, or recoverable retries; and keep task-level scorers, audit logs, and review queues beside schema checks. Extra retries, graph nodes, and optimization loops consume tokens and latency and should be justified by measured lifts under a common call budget. Empty traces or framework-native ``success'' flags without boundary revalidation are weak evidence for production readiness.

Practitioner guidance remains partially hypothesis-level for mechanism choice, but the confirmatory evidence already supports layered caution: schema acceptance is not task correctness; SDK references are necessary baselines; and framework adoption should be justified by measured gains on the team's contracts. Section~\ref{sec:business} develops the corresponding business-process mapping, stack selection narrative, and operating-model reading for analytics and risk owners.

\section{Business Implications}
\label{sec:business}

The confirmatory matrix is a technical evaluation, but the tasks are business interfaces. Documents become records, questions become queries, policies become answers, and requests become routing decisions. Once an LLM output crosses that boundary, the organization is no longer debating model eloquence; it is deciding whether a value may enter a ledger, warehouse, case system, or approval queue. This section translates the evidence into deployment guidance for analytics, operations, and risk owners without upgrading system-configuration contrasts into vendor rankings.

\subsection{Where Schema-Validated Workflows Show Up}

Table~\ref{tab:business-tasks} maps each benchmark family to a familiar enterprise process. The mapping is deliberate: these are not toy prompts, but the kinds of interfaces where a wrong field can post an invoice, a wrong predicate can leak a join, or a wrong escalation flag can leave a customer request unreviewed.

\begin{table}[t]
\centering
\small
\begin{tabularx}{\textwidth}{P{0.18\textwidth}P{0.28\textwidth}Y}
\toprule
Task family & Business process & What ``good'' must mean operationally \\
\midrule
Extraction & Accounts payable / intake automation & Fields are present \emph{and} match the source document before posting. \\
SQL generation & Self-serve analytics & Query executes under least privilege and returns the intended grain. \\
Summarization & Management reporting & Claims are supported by the report; fluency is not enough. \\
Compliance QA & Policy / controls advisory & Answers cite the governing text and abstain when unsupported. \\
Workflow decision & Tier-1 support / case routing & Tool choice and human-review flags match operating procedure. \\
\bottomrule
\end{tabularx}
\caption{Benchmark task families as business interfaces. Schema validity is an admission control for these processes, not a substitute for process controls.}
\label{tab:business-tasks}
\end{table}

In all five cases the expensive failure is rarely an unparsable string. It is a well-formed object that looks operationally complete and therefore travels farther into the enterprise than a crash ever would. That is why schema dashboards can improve while process risk does not: the system has gotten better at admitting candidates, not necessarily better at admitting truth.

\subsection{What the Confirmatory Evidence Means for Buyers and Builders}

Under the claim boundary in Section~\ref{sec:research_questions}, three business readings follow from the confirmatory run.

Parseable is not postable. Overall schema validity of 0.684 with mean automated quality of 0.638, and frequent metric flags among schema-valid extraction and compliance rows, means finance and risk teams should not treat ``JSON mode succeeded'' as an approval to write to ERP, data warehouse, or case systems. Recent validity--correctness measurements make the same warning in different language when they urge separate reporting of schema validity and answer or executable accuracy \citep{ray2026constrainttax}. Application schemas reduce malformed writes. Task checks and review queues still decide whether a record is trustworthy.

Integration paths also differ under the same model. Near-ceiling structural performance for PydanticAI, LangGraph, and DSPy versus materially lower performance for the plain SDK, Agno, and CrewAI conditions in this package state shows that the binding layer between model and contract is itself a business risk factor. Procurement language that stops at ``supports structured output'' is therefore incomplete. Architecture reviews should ask how schemas are passed, how retries are counted, whether acceptance is revalidated at the application boundary, and whether an SDK baseline exists against which framework ceremony can be judged.

Baselines prevent false confidence. Keeping an SDK reference is a governance practice as much as a research design. If a framework appears to win only because the baseline never enforced the same schema, the organization has bought ceremony rather than reliability. Conversely, if a typed or orchestrated path measurably reduces rejects, repair loops, or review burden on the team's own contracts, the framework cost has a business justification that can be stated in operating terms rather than in vendor preference.

\subsection{Choosing a Stack Without Buying a Slogan}

Stack selection should begin with the contract and the system of record, not with a framework shortlist. An invoice line, SQL result set, policy answer with a citation span, or routing decision with a review flag is a business object first. Writing that schema---and naming where a rejected or uncertain object goes---clarifies whether the immediate need is shape control, process state, or metric-driven prompt search.

When the process is essentially one validated object and failures can fail closed, provider structured output plus an SDK path is often the right first experiment. It establishes what the model and provider already deliver before any orchestration tax is paid. Application validation and typed agents become valuable when silent shape drift is costly: posting, entitlements, and downstream joins are exactly the settings where fail-closed validation and explicit repair beat logging-only checks. Orchestration graphs earn their keep when the \emph{process} needs state---repair loops, tool execution, approvals, or resumable work---not merely when structured extraction feels hard. Schemas address that class of problem first; graphs address durable control flow.

Optimization loops such as DSPy are a second cycle around a contract that already has a stable metric and labeled examples. Without those, optimization amplifies the wrong objective with great efficiency. Multi-agent wrappers are optional composition after the contract boundary is proven. Role and task frameworks can help divide labor, but they must still survive the same schema and task gates that a single typed call would face. Across all of these choices, human review should be budgeted by task risk. In this study, extraction and compliance QA warranted more metric flags than summarization; high-impact writes should keep sampling review even when schema dashboards look green.

\subsection{Operating Model: Controls, Cost, and Accountability}

Business value appears when technical layers map cleanly onto operating controls. Provider constraints and application schemas are preventive controls: they decide what may enter the pipeline. Task scorers, audit packets, and human review are detective controls: they decide what may be trusted after entry. Retries and routing are corrective controls: they decide how the organization recovers when prevention fails. Confusing these roles produces a familiar failure mode in which dashboards report structural success while incorrect amounts or unsupported policy claims still move work downstream.

Cost follows the same map. Extra provider calls, graph nodes, and optimization runs are operating expenses, not free reliability. They should be authorized against measured lifts on reject rates, repair time, or review burden under a declared call budget---the same discipline used in the frozen analysis plan. Empty traces and framework-native success flags without boundary revalidation are weak evidence for audit or change-control purposes, because they do not reconstruct what was accepted, why it was accepted, or what business check still remained open.

Accountability should stay with the process owner. The benchmark does not certify unsupervised financial, compliance, or customer decisions. Schema-validated workflows are safest when they accelerate drafting and structuring work that still sits inside existing approval, logging, and least-privilege regimes. In that framing, the technology is an assistant to a controlled process rather than a substitute for one.

\subsection{Limits of the Business Reading}

These implications inherit the study's limits: synthetic tasks, a primary model family, package-sensitive adapters, incomplete mechanism ablations, and unreviewed human-audit packets for semantic-error prevalence. They are decision aids for designing controls and evaluation programs. They are not a league table of vendors, and they are not a license to remove human oversight from high-impact workflows.

\section{Threats to Validity and Limitations}
\label{sec:threats}

Table~\ref{tab:limitations-summary} consolidates the main limitations of the confirmatory study and its ablations.

\begin{table}[t]
\centering
\small
\begin{tabularx}{\textwidth}{P{0.20\textwidth}Y Y}
\toprule
Threat & Consequence for this paper & Mitigation in the confirmatory design \\
\midrule
Construct-limited scores & Automated flags are not human-confirmed semantic error rates. & Deterministic scorers plus blinded audit packets; flags reported as candidates. \\
Implementation mismatch & The LCD matrix uses only a subset of each library's advertised mechanisms. & Labeled feature-using ablations for RQ4; attribute results to concrete configurations. \\
Adapter fragility & CrewAI/Agno surfaces can change with package versions. & Pin versions in run provenance; retain SDK baseline. \\
Synthetic data & Clean templates may not represent document noise or production tools. & Stratify harder synthetic cases; report licensed public-data extensions separately. \\
Single model and time period & Provider behavior and library integrations can change. & Pin versions; optional secondary-model ablation subset. \\
Incomplete mechanism coverage & Retry/validator/typed-deps ablations do not exhaust RQ4. & Expand ablations before mechanism-level product claims. \\
\bottomrule
\end{tabularx}
\caption{Primary validity threats, their effect on confirmatory claims, and mitigations.}
\label{tab:limitations-summary}
\end{table}

Internal validity is limited by system-level attribution. A condition includes provider settings, framework APIs, adapter code, prompts, retry logic, and scorer behavior. Researcher familiarity may also make one adapter more idiomatic than another. Confirmatory claims therefore refer to versioned implementation conditions and use feature ablations rather than treating a package name as the causal intervention.

Construct validity is limited by the automated metrics. Exact or normalized field matching is suitable for deterministic synthetic fields, but policy entailment, summary factuality, and workflow completion require richer evidence. LLM judges can introduce position, verbosity, and self-preference biases \citep{zheng2023judge}; they should triage cases rather than replace deterministic checks and blinded human adjudication. The current human-audit packet is unreviewed.

Conclusion validity for the confirmatory matrix rests on instance-clustered estimands, distinct-instance effective sample sizes, frozen equivalence margins, and Holm multiplicity as committed in \texttt{config/analysis\_plan.json}. Ablation pilots with smaller per-cell $n$ support directional mechanism evidence, not the same precision as the primary matrix.

External validity is limited to synthetic business tasks, the primary model family, and the package state recorded in eligible run provenance. Results do not generalize to long-lived agents, human-shared control, production databases, regulated decisions, or adversarial inputs. Schema validation also does not address prompt injection, unsafe authorization, data leakage, or supply-chain risk \citep{owaspLLMTop102025}. The benchmark evaluates one application boundary; it is not a security or deployment certification.

\section{Artifact, Data, and Ethical Considerations}
\label{sec:artifact_ethics}

\subsection{Availability and Provenance}

The public code artifact is available at
\url{https://github.com/ULDataProfessor/schema-validated-llm-workflows}.
It contains the benchmark code, synthetic-data generator, task schemas, prompts, six adapters, deterministic scorers, no-API tests, frozen analysis settings, and protocol documentation. The arXiv source archive rebuilds this manuscript and is not a substitute for that repository. Live result directories remain local by default (they are gitignored) and may be redistributed separately when redacted; cloning the repository alone is therefore sufficient to rerun the pipeline, but not to replay a specific archived result bundle without that bundle.

Eligible confirmatory and ablation runs record provenance fields in \texttt{meta.json} (command arguments, timestamps, package versions, git state, and configuration hashes). Claims that depend on a cited run should reference that run identifier and its archived aggregate records.

\subsection{Data and Privacy}

The current tasks use synthetic invoices, complaints, reports, policies, SQL databases, and support requests. They do not intentionally contain real customers, employees, accounts, or confidential organizational facts. No human-subject interaction was performed, and the empty human-audit worksheet is not human-participant data.

Model traces can contain prompts, source documents, tool arguments, and generated text. A public release should retain metric-critical fields and cryptographic links to redacted payloads when full text cannot be redistributed. Redaction must be explicit: a reserialized parsed object is not a substitute for a raw provider response unless it is labeled as such. Future public-data extensions require dataset-specific licensing, privacy review, acquisition dates, and reproducible conversion scripts.

\subsection{Responsible Interpretation}

The benchmark does not certify any condition for unsupervised financial, compliance, or operational use. Schema-valid fields can appear authoritative after entering a database even when they are unsupported. High-impact deployments should combine runtime schemas with domain validation, least-privilege tools, abstention, review queues, audit logs, and monitoring. Full-matrix API calls also carry financial and environmental costs; token and call counts should be reported so that additional runs are proportionate to their evidentiary value.

\section{Conclusion}
\label{sec:conclusion}

Schema-validated LLM workflows need two measurements that practice often collapses into one: whether an output satisfies a runtime contract, and whether that accepted value is correct for the business task. This paper contributes a protocol that keeps those layers separable, a public executable artifact across six Python implementation conditions, and a confirmatory 5{,}400-row live-API matrix under a frozen analysis plan powered on distinct instances.

The confirmatory evidence supports the working hypotheses at the system-configuration level. Structural validity and task quality diverge (H1), especially on extraction and compliance. Binding paths are heterogeneous under a shared LCD contract (H2): PydanticAI, LangGraph, and DSPy remain near ceiling while the plain SDK, Agno, and CrewAI conditions do not. Gaps are interactive (H3): summarization compresses differences, while particular cells---notably CrewAI on workflow routing---fail far beyond additive expectations. Practically, teams should treat provider-native JSON modes, application schemas, and task evaluators as complementary controls; keep an SDK baseline; and justify framework adoption with measured gains on their own contracts.

Labeled ablations begin the RQ4 program (retry budgets; validators and typed dependencies on PydanticAI). Completed human adjudication of metric-flagged candidates, broader maintainability or stress claims, and full mechanism attribution remain open. For business deployment, Section~\ref{sec:business} frames the same evidence as process controls: schemas as admission gates, task checks and review as detective controls, and framework choice as a measured investment rather than a default.

\bibliographystyle{plainnat}
\bibliography{references}

\appendix
\section{Reproducibility Appendix}
\label{sec:reproducibility}

\subsection{Repository Layout}

The public code artifact is
\url{https://github.com/ULDataProfessor/schema-validated-llm-workflows}.
Manuscript sources are maintained separately from that repository.

\begin{lstlisting}[style=yamlstyle,caption={Relevant public-artifact layout.},label={lst:repo-layout}]
.
|-- docs/                     # Protocol and audit documentation
|-- scripts/cloud/            # Optional cloud provisioning helpers
`-- benchmark/
    |-- config/               # Model, trial, ablation, stress settings
    |-- data/schemas/         # Shared task schemas
    |-- data/synthetic/       # Generated JSONL (ignored by Git by default)
    |-- tasks/                # Shared prompt construction
    |-- frameworks/           # Six implementation adapters
    |-- eval/                 # Scoring, audit, and reporting code
    |-- scripts/              # Generation and run commands
    `-- results/              # Run bundles (ignored by Git by default)
\end{lstlisting}

The synthetic JSONL files and result directories are deliberately ignored by Git because live outputs may be large or sensitive. A reproducible result release must therefore archive them explicitly; cloning the code repository alone is sufficient to regenerate data and rerun the pipeline, but not to replay a cited result directory without that directory.

\subsection{Environment and No-API Checks}

\begin{lstlisting}[style=yamlstyle,caption={Local setup and no-API validation.},label={lst:setup-checks}]
cd benchmark
python -m venv .venv
source .venv/bin/activate
pip install -e ".[all]"
python scripts/generate_synthetic_data.py
python -m pytest tests
python scripts/plan_matrix.py
python scripts/check_readiness.py --dry-run
python scripts/run_benchmark.py --all --pilot --dry-run
\end{lstlisting}

The project specifies dependency lower bounds rather than a lock file. Eligible runs write an environment snapshot to \texttt{meta.json}; bit-for-bit recreation additionally benefits from an exported lock or fully resolved package list.

\subsection{Live Run Workflow}

\begin{lstlisting}[style=yamlstyle,caption={Pilot and confirmatory run entry points.},label={lst:live-commands}]
# After provider keys and optional imports pass readiness checks:
python scripts/check_readiness.py --pilot
python scripts/run_benchmark.py --all --pilot --run-id <pilot_id>

# Launch only after the hardening gate in Section 6 passes:
python scripts/check_readiness.py
python scripts/run_benchmark.py --all --all-seeds --run-id <full_id>

# Resume uses the same id and validates compatible metadata:
python scripts/run_benchmark.py --all --all-seeds \
  --run-id <full_id> --resume
\end{lstlisting}

The runner writes each completed row before the next call. Current metadata capture records command arguments, timestamps, Python and platform versions, selected package versions, git state, and hashes for committed configuration files. An eligible release must additionally record prompt, schema, scorer, and adapter versions plus the attempt-level fields specified in Section~\ref{sec:benchmark}.

\subsection{Result Bundle}

\begin{lstlisting}[style=yamlstyle,caption={Minimum result-bundle structure.},label={lst:result-contract}]
benchmark/results/<run_id>/
|-- meta.json
|-- trials.jsonl
|-- summary.json
|-- tables.tex
|-- result_card.md
|-- artifact_audit.{md,json}
|-- claim_records.{md,json}
|-- failure_examples.{md,json}
|-- human_audit.{md,csv,json}
|-- human_audit_summary.{md,json}
|-- cost_analysis.{md,json}
`-- paired_<comparison>.json       # required for comparative claims
\end{lstlisting}

\texttt{trials.jsonl} is the authoritative metric source; summaries and tables are derived. A public bundle may replace sensitive raw payloads with encrypted files, redacted excerpts, or hashes, but it must state what was removed and retain enough structured evidence to recompute the reported diagnostics. Historical rows that contain a reserialized parsed object in \texttt{raw\_output} must be labeled as such.

\subsection{Paper and Release Checks}

\begin{lstlisting}[style=yamlstyle,caption={Release checks from the repository root.},label={lst:release-checks}]
python scripts/check_release_readiness.py \
  --json-output release_readiness.json \
  --markdown-output release_readiness.md

cd paper
make check
make package
\end{lstlisting}

The release checker runs no-API benchmark tests, manifest validation, the LaTeX build, bibliography and cross-reference checks, and pilot claim-scope checks. The paper packager follows \LaTeX{} input dependencies so generated tables are included in the arXiv source archive. These checks establish internal consistency, not empirical correctness.

For a public release, the extracted manuscript source archive should compile in a clean directory, and any redistributed result bundle should be checked for absolute local paths, usernames in archive headers, secrets, provider-restricted content, and missing generated data. The public code URL for this project is \url{https://github.com/ULDataProfessor/schema-validated-llm-workflows}.

\section{Implementation and Ablation Patterns}
\label{sec:implementation_patterns}

The confirmatory LCD matrix uses the structured-output mechanisms summarized in Table~\ref{tab:conditions}. This appendix records the stronger implementation boundaries required for mechanism attribution; the examples are design specifications for feature-using ablations unless a labeled run activates them.

\subsection{Independent Acceptance Boundary}

Every adapter should return a candidate plus native evidence. The benchmark, rather than the adapter, then performs the final shared-schema check:

\begin{lstlisting}[style=pythonstyle,caption={Independent benchmark-boundary validation.},label={lst:independent-validation}]
def accept_candidate(task, native_result, schema):
    candidate = native_result.candidate
    try:
        value = schema.model_validate(candidate)
        benchmark_valid = True
        validation_error = None
    except ValidationError as exc:
        value = None
        benchmark_valid = False
        validation_error = str(exc)

    return TrialRecord(
        task=task.name,
        native_accepted=native_result.accepted,
        benchmark_valid=benchmark_valid,
        provider_attempts=native_result.provider_attempts,
        retry_events=native_result.retry_events,
        request_hash=native_result.request_hash,
        parsed_output=value,
        validation_error=validation_error,
    )
\end{lstlisting}

This boundary preserves three facts that must not be conflated: whether the provider accepted the request, whether the library returned a typed object, and whether the shared benchmark schema accepted the candidate. Attempt counts must be derived from observed provider calls rather than a runner-local counter.

\subsection{Feature-Using PydanticAI Condition}

PydanticAI's application-level value cannot be tested by \texttt{output\_type} alone. A feature-using ablation should add a domain validator whose effect is observable and scored. For SQL, the validator can reject non-read-only queries and run a safe check on an isolated database:

\begin{lstlisting}[style=pythonstyle,caption={Illustrative PydanticAI domain-validator condition.},label={lst:pydantic-validator}]
from pydantic_ai import Agent, ModelRetry, RunContext

agent = Agent(
    model,
    output_type=SQLGenerationResult,
    deps_type=ReadOnlyDatabaseFactory,
    output_retries=common_repair_budget,
)

@agent.output_validator
async def validate_sql(
    ctx: RunContext[ReadOnlyDatabaseFactory],
    result: SQLGenerationResult,
) -> SQLGenerationResult:
    if not is_single_read_only_select(result.sql):
        raise ModelRetry("Return one read-only SELECT statement.")
    with ctx.deps.fresh_database() as connection:
        if not explain_succeeds(connection, result.sql):
            raise ModelRetry("The query does not compile on the supplied schema.")
    return result
\end{lstlisting}

The comparison should include the same provider mode with this validator disabled. A reported gain must identify whether it comes from rejection, successful repair, or safer review routing and must include the additional calls and latency \citep{pydanticaiOutput2026,pydanticaiDeps2026}.

\subsection{Mechanism-Specific Conditions}

Table~\ref{tab:mechanism-evidence} defines the minimum active condition and trace evidence for each mechanism-level comparison. These are requirements for RQ4 ablations, not automatic properties of the LCD confirmatory matrix.

\begin{table}[htbp]
\centering
\small
\begin{tabularx}{\textwidth}{P{0.19\textwidth}P{0.30\textwidth}Y}
\toprule
Mechanism & Minimum active condition & Required evidence \\
\midrule
Provider enforcement & Strict schema mode versus relaxed generation with the same final validator & Exact request mode, provider acceptance, parse result, and independent validity \\
Retry feedback & Common call budget with and without validation feedback & Attempt count, feedback text or hash, retry cause, final status, tokens, and latency \\
Typed dependencies/tools & Versioned resource passed through a typed boundary and an executed tool & Dependency version, validated arguments, authorization, tool result, and final disposition \\
Graph state & Multiple nodes, persisted state, and a review or recovery interrupt & State transitions, checkpoint/restart evidence, approval decision, and executed outcome \\
DSPy optimization & Frozen train/development/test split and declared optimizer & Optimizer configuration, development calls and cost, compiled program, and held-out results \\
\bottomrule
\end{tabularx}
\caption{Evidence required to attribute an outcome to a concrete mechanism rather than a framework label.}
\label{tab:mechanism-evidence}
\end{table}

\subsection{Adapter Surface Diagnostics}

Table~\ref{tab:current-complexity} reports static counts from the current runner directories. The counter excludes tests and counts nonblank Python lines plus visible framework-specific API markers. These numbers describe adapter surface area; they do not measure developer effort or maintainability and are not used for ranking.

\begin{table}[htbp]
\centering
\small
\begin{tabularx}{\textwidth}{lrrrY}
\toprule
Condition & Files & LOC & Markers & Optional dependencies \\
\midrule
PydanticAI & 1 & 162 & 2 & \texttt{pydantic-ai} \\
Plain SDK & 1 & 146 & 7 & none beyond shared dependencies \\
LangGraph & 1 & 193 & 6 & \texttt{langgraph}, \texttt{langchain-openai} \\
Agno & 1 & 142 & 3 & \texttt{agno} \\
DSPy & 1 & 217 & 6 & \texttt{dspy} \\
CrewAI & 1 & 237 & 6 & \texttt{crewai} \\
\midrule
Total & 6 & 1,097 & 30 & -- \\
\bottomrule
\end{tabularx}
\caption{Static surface of the current implementation adapters. LOC and marker counts are descriptive artifact diagnostics, not maintainability ratings.}
\label{tab:current-complexity}
\end{table}

\subsection{Schema-Change Exercise}

Maintainability would be evaluated with a versioned change request, not line count alone. A representative exercise would add a required evidence span and explicit review disposition to the compliance output. Each implementation would receive the same change specification and failing tests. The resulting artifact would record files touched, elapsed implementation and debugging time, test failures, backward-compatibility decisions, and a blinded reviewer rating. No maintainability result is claimed in this manuscript.

\end{document}